# Effect of dilution of both A- and B- sites on the multiferroic properties of spinal Mott insulators

Prashant Shahi[1], R. Singh[1], Shiv Kumar[2], A.Tiwari[4], A.Tripathi[4], J. Saha[5], S. Patnaik[5], A. K. Ghosh[2] and Sandip Chatterjee[1,*]

[1]Department of Physics, Indian Institute of Technology (Banaras Hindu University), Varanasi-221 005, India

[2]Department of Physics, Banaras Hindu University, Varanasi-221 005, India

[3]Department of Physics, National institute of Technology Durgapur, India

[4]Department of Physical Sciences, School of Chemical and Physical Sciences, Sikkim University, Sikkim, India

[5]School of Physical Sciences, Jawaharlal Nehru University, New Delhi-110079, India

## Abstract

The structural, magnetic, electrical and transport properties of $FeV_2O_4$, by doping Li and Cr ions respectively in A and B sites, have been studied. Dilution of A-site by Li doping increases the ferri-magnetic ordering temperature and decreases the ferroelectric transition temperature. This also decreases the V-V distances which in effect increases the A-V coupling. This increased A-V coupling dominates over the decrease in A-V coupling due to doping of non-magnetic Li. On the other hand, Cr doping increases the ferri-magnetic ordering temperature but does not alter the ferroelectric transition temperature which is due to the fact that the polarization origin to the presence of almost non-substituted regions.

*Corresponding author *e-mail id*: schatterji.app@iitbhu.ac.in
*Ph.No. & Fax No.: +91-5426701913*

## Introduction

Transition- metal- oxide (TMO) spinals are becoming the centre of attraction with significant quality of complex interaction among charge, spin and orbital degrees of freedom which gives it some interesting properties. Importantly, vanadium spinel oxides $AV_2O_4$ (A =$Fe^{2+}$, $Mn^{2+}$, $Co^{2+}$, $Zn^{2+}$, $Mg^{2+}$), where $A^{2+}$ and $V^{3+}$ ions occupy the tetrahedral (A site) and octahedral (B site) sites, respectively, have two 3d electrons in the triply degenerate $t_{2g}$ states at $V^{3+}$ site. This whole arrangement is becoming the centre of attention because of their fascinating magnetic and orbital order.[1-7] When A site is replaced by some non magnetic ion [Li, Zn, Mg, Cd] it shows many interesting properties. Likewise $ZnV_2O_4$ goes from cubic to tetragonal state at T=50 K and orders antiferromagnetically at T= 42 K[8] and $LiV_2O_4$ shows metallic behavior and it is a first transition metal oxide which shows heavy fermion behavior and remains in cubic phase in its whole temperature range.[9-11] On the other hand, when A site is replaced by Magnetic ions, different properties emerges. In this way the $MnV_2O_4$ exempts magnetic phase transition from the collinear ferrimagnetic order to the non-collinear canted one at 53 K with associations to the anti-ferro-orbital order of vanadium $t_{2g}$ orbitals.[3,12] Whereas, $CoV_2O_4$ exhibits two magnetic phase transitions at 142 K and 59 K without any structural phase transitions.[13,14]

Moreover, $FeV_2O_4$ is a unique compound among such spinel vandium oxides comprising both $Fe^{2+}$ and $V^{3+}$ ions with orbital degrees of freedom; $Fe^{2+}$ ion at the tetrahedral site (A-site) having three 3d electrons in the doubly degenerate $e_g$ states. Very recent development shows structural and magnetic properties of the spinal $FeV_2O_4$ [11] exhibiting successive structural transitions from cubic to compressed tetragonal with the lattice constants of $c <a$ at $Ts_1 \sim 140$ K which is due to the cooperative Jahn–Teller effect of $FeO_4$ and from tetragonal to orthorhombic transition accompanied by a ferrimagnetic transition at $T_{s2} \sim 110$ K, and from orthorhombic to elongated tetragonal with $c >a$ at $T_{s3} \sim 60$ K with decreasing temperature for polycrystalline samples. It has also been reported that the Jahn–Teller effects and the relativistic spin–orbital coupling play an important role in the determination of the orbital states of $Fe^{2+}$ and $V^{3+}$ ions at low temperatures which were concluded with the help of single crystal x ray diffraction experiments.[6] Recent reports of NMR and neutron diffraction of $FeV_2O_4$ indicate that its structure is changing to non-collinear ferrimagnetic state at 60 K from collinear state, where the $V^{3+}$ moments become mounted along the {111} directions.[7,15] This latter transition is marked by a step in magnetization, a peak in heat capacity, an anomaly in the dielectric constant, and the appearance of polarization. It was found that the application of a magnetic field shifts all these signatures associated to $T_{s3}$ to higher temperatures, while it also clearly affects the value of the polarization, revealing a significant magneto electric coupling. It is suggested that the presence of canted spins in the triangular structure below $T_{s3}$ could be responsible for the appearance of ferroelectricity.[16]

The $AV_2O_4$ system also approaches the itinerant-electron limit with decreasing V-V separation.[17-18] The predicted critical separation for metallic behavior is $R_c = 2.94$Å.[19] A Recent study on $FeV_2O_4$ and $CoV_2O_4$ shows that with increasing pressure the V-V separation decreases and due to which there is a delocalization of charge carriers in $FeV_2O_4$ and it induces metallic behavior in $CoV_2O_4$.[14] The same effect is also shown by chemical pressure by doping Co at the site of $MnV_2O_4$.[20] Recently it is shown that in $FeV_2O4$, $CoV_2O_4$ and $MnV_2O_4$ the magnetic transition temperature suppressed and activation energy decreases as $Zn^{2+}$ (non magnetic ion) doped at the A site.[21-23] Furthermore, $Li^{1+}$ is also a non magnetic ion and its size is comparable to Mn, Zn, Fe and Co but $Li^{1+}$ has no 3d electron in their outer shell unlike Fe and Co. Therefore, it will be interesting to investigate the magnetic and transport properties by doping Li in $CoV_2O_4$ and $FeV_2O_4$ at the A site.

Furthermore, soft magnetic materials are center to nearly every aspect of modern electrical and electronics technology because of their ability to concentrate and to shape magnetic flux with great efficiency. The most important characteristics desired for essentially all soft magnetic applications are high saturation induction, high permeability, low coercivity, and low core loss. In this regard these materials are highly important from the application point of view.

In this paper, we have investigated magnetic, electrical polarization and transport properties of $Li_xFe_{1-x}V_2O_4$ (0≤x≤0.1) and found that with increasing the Li and Cr content at Fe (A) and V (B) sites the Ferrimagnetic transition temperature increases, electric polarization temperature remains same with Cr doping and decreases for Li doping and the system moves towards iterant electron limit due to decrease in the V-V distance with Li doping and moves towards more localized behaviour with Cr doping .

## Experiment

The polycrystalline Fe(V$_{1-x}$Cr$_x$)$_2$O$_4$ (0≤x≤0.05) and Li$_x$Fe$_{1-x}$V$_2$O$_4$(0≤x≤0.1) samples used in this study were prepared by solid state reaction method. Appropriate ratio of Li$_3$VO$_4$, Fe, Fe$_2$O$_3$, Cr$_2$O$_3$, V$_2$O$_3$ and V$_2$O$_5$ were grounded thoroughly and pressed into pellets. The pellets were sealed in evacuated quartz tube and heated at 1050$^o$C for 40 hours for Li$_x$Fe$_{1-x}$V$_2$O$_4$ (0≤x≤0.1). The X-ray powder diffraction experiment has been performed using Rigaku Mini Flex II DEXTOP X-ray Diffractometer with Cu-kα radiation. Magnetic measurement was done using MPMS SQUID (Quantum Design) magnetometer with the bulk samples. Ac-susceptibility measurement were done using lock in amplifier SRS830 by homemade setup and standardized with YBCO superconducting sample. The electric polarization (P) as a function of temperature was determined by integrating the pyroelectric current measured on warming at a rate of 5 K/min with a Keithley 6517A electrometer then cooled with a poling electric field of 640 kV/m applied from 90 K to 6 K. At 6 K, the poling electric field was removed and the electrodes were shortened for 1h to reduce the possible contribution from the detrapped charges. Resistivity measurements have been done using four probe method.

## Results and Discussions

Figure 1 shows the X-ray diffraction (XRD) pattern for different Li and Cr doped samples. All peaks are indexed with Fd-3m space group indicating our samples are of pure phase. Inset of the figure 1 shows variation of lattice parameter with Li and Cr Doping, obtained from the Rietveld refinement of the XRD data. It is observed that with increasing the Li concentration at the Fe site the lattice parameters decrease linearly following the Vegard's law. The same trend is followed by Cr doping also. The ionic size of Li is 0.73Å which is smaller than Fe (0.77Å) and V$^{3+}$ ionic size is 0.78 Å is greater then Cr$^{3+}$ (0.75 Å) ionic radius. As a matter of fact, the lattice parameters decrease with Li and Cr doping. In case of Li doping the lattice parameter decreases very sharply with respect to Cr doping which might be due to the fact that in Li no 3d electrons present in outer shell due to which the Coulomb repulsion decreases between Li/Fe and oxygen 2p electrons. The parameters obtained from Reitveld refinement are given in the Table 1. Figure 2 shows the variation of ZFC and FC magnetization with temperature for Li$_x$Fe$_{1-x}$V$_2$O$_4$ (0≤x≤1) and Fe(V$_{1-x}$Cr$_x$)$_2$O$_4$ (0≤x≤0.05). For FeV$_2$O$_4$ (Fig. 2(a)), with decrease of temperature, a transition from paramagnetic (PM) to ferrimagnetic (FI) phase occurs below Tc. ZFC magnetization bifurcates from FC magnetization around 112 K, then decreases with further decreasing temperature. A small rise in FC curve exists at a lower temperature Tp, accompanied by a slight drop in ZFC magnetization. This magnetization behavior is consistent with that already reported.[24] The values of Tc and Tp determined from the dM/dT vs. T curve of the FC magnetization, are 112K and 62K, respectively. With Li 10% doping the $T_C$ increase to 126 K but the $T_P$ disappears whereas for 5% Cr doping $T_C$ increaess to 120 K but $T_P$ remains unchanged. We have also measured ac-susceptibility at 100 Hz for $Li_xFe_{1-x}V_2O_4$ ($0 \leq x \leq 1$) which is shown in figure 3. Similar behavior is observed that with increasing Li content at A site the ferrimagnetic transition temperatures increase. This might be due to the fact that shrinkage of the lattice parameter with Li and Cr doping, increases the exchange interaction between the $A^{2+}$ and $V^{3+}$ through oxygen which enhances the ferrimagnetic ordering temperature.

Figure 4 shows the M(H) curve at 2K for undoped and Li and Cr doped samples. For the undoped sample a jump at ~0T and another jump ~1.25T are observed which are consistent with that reported by Nishihara et. Al.[24] But when Li is doped the jumps disappear. On the other hand when Cr is doped the 0T jump shifts to 1.0T and the 1.25T jump shifts to 1.3 T and an extra jump is observed at ~2.5T. In their paper.[24] Nishihara et al explained the 1.2T jump in $FeV_2O_4$ as the avalanche behavior.[25] But the origin of jumps in Cr doped sample are not yet clear. It might be due to some Martensitic transition.[26] Anyway, it deserves further study. On the other hand, the saturated moment, estimated from figure 4, decreases upon Li doping and increases for Cr doping. Since Li ions are non-magnetic and Cr ions are more magnetic than V ions and it is known that V spins tend to align anti-parallel to each other when the coupling between A and B sub-lattices is absent. Therefore, the moment of the V sub-lattice also decreases with increasing Li content and increases for Cr content. It is important to compare these with the earlier studies on $Zn_xMn_{1-x}V_2O_4$ [23] and $Co_{1-x}Zn_xV_2O_4$ [22] because Zn(0.74Å) is also a non magnetic ion. Moreover, the sizes of Li (0.73Å), Co (0.72Å) and Zn (0.74Å) are comparable. In case of $Zn_xMn_{1-x}V_2O_4$ it is found that as the Zn content increases the magnetic transition temperature decreases which is different than that in Li doped samples. With Zn doping two cases arises, the lattice parameters decrease with increasing Zn content due to which the V-V distance decreases so that coupling between A and V sites increase which tries to increase the magnetic transition temperature. But doping of non magnetic ion at A site decreases the A-V coupling due to which V-V spins try to align anti-parallel. As a result, the magnetic transition is suppressed.[23]

Figure 5 shows the temperature dependent electrical polarization for the $FeV_2O_4$, $Li_{0.05}Fe_{0.95}V_2O_4$ and $Fe(Cr_{0.05}V_{0.95})_2O_4$ samples. The electrical polarization (P) appears at about $T_p$ = 61 K for $FeV_2O_4$ and 38 K for $Li_{0.05}Fe_{0.95}V_2O_4$ and approaching a saturated value with decreasing temperature. For 10% Li doping no polarization is observed consistent with the magnetization behavior. With Li doping, also the onset temperature decreases. This result emphasizes that this substituted $FeV_2O_4$ remains a multiferroic until dilution becomes too detrimental to the magnetic ordering. But with Cr doping the onset temperature remains same. In contrast, for the Cr doped samples ($Fe(Cr_{0.05}V_{0.95})_2O_4$), the $P(T)$ curves reveal unchanged $T_P$ values as $x$ increases. The temperature, $T_P$ *(ferroelectric transition temperature)*, is not changing with $Cr$ content, but the $P$ value decreases with increase of $Cr$ content. It is thus reasonable to ascribe the polarization origin to the presence of almost non-substituted regions (with compositions near $FeV_2O_4$ and whose amount decreases with increasing $x$). From above discussion it is clear that Li doping suppresses the collinear to non collinear state but Cr doping does not change the collinear state to non collinear state, which might be due to the fact that with Li doping the coupling between V-V decreases due to the presence of $V^{3+}$ and $V^{4+}$ at the octahedral site but Cr 5% doping has no effect on V-V coupling which in effect does not change the onset temperature.

Figure 6(a) shows the variation of resistivity with temperature for all the samples and Figure 6(b) shows the ln ρ vs 1000/T for $Fe_{1-x}Li_xV_2O_4$ [where x=0 and 0.1] and $Fe(Cr_{0.05}V_{0.95})_2O_4$ [where x=0.05]. As $FeV_2O_4$ belongs to Mott Insulator regime therefore from the figure it can be mentioned that with increasing Li content at the A site the system moves towards itinerant electron side along with the decrease of V-V distance. The 3*d* electrons in Cr ions are well localized. Therefore, in the Cr-doped samples the electric resistivity shows an insulating behavior.

## Conclusion

The structural, magnetic, electrical and transport properties of $FeV_2O_4$, by diluting A-site by Li and B-site by Cr, have been studied. It is observed that increasing the Li content at A site the V-V distance decreases and due to that ferrimagnetic ordering temperature increases while the ferroelectric transition temperature decreases and the whole system moves towards the itenerant electron behavior. In the case of Cr doping the magnetic transition temperature increases but there is no effect on the ferroelectric transition temperature and the system moves towards more inside the Mott insulating region. So by tuning the V-V distance either by external pressure or chemical pressure we can tune the magnetic, ferroelectric and transport properties or more specifically, the multiferroic properties of Mott-insulating $FeV_2O_4$ which is very important from the application point of view.

**Acknowledgement**

SC is grateful to DST, India (Grant No.: SR/S2/CMP-26/2008), CSIR, India (Grant No.: 03(1142)/09/EMR-II) and BRNS, DAE, India ((Grant No.: 2013/37P/43/BRNS) for providing financial support. PS is grateful to CSIR, India for providing financial support.

**Table 1**. Structural parameters (lattice parameters, bond lengths) of $Fe_{1-x}Li_xV_2O_4$ ($0<x<0.1$)) samples obtained from Rietveld refinement of X-ray diffraction data

| Sample Name | a(Å) | d(V-V)(A$^0$) |
|---|---|---|
| $FeV_2O_4$ | 8.4517 | 2.9881 |
| $Fe_{0.95}Li_{0.05}V_2O_4$ | 8.4429 | 2.9850 |
| $Fe_{0.9}Li_{0.1}V_2O_4$ | 8.4341 | 2.9819 |
| $Fe(Cr_{0.05}V_{0.95})_2O_4$ | 8.4485 | 2.9870 |

## Figure Captions:

1. X-ray diffraction pattern with Reitveld refinement for Li and Cr doped $FeV_2O_4$ samples at 300K. The inset shows the variation of lattice parameters with Li and Cr concentration

2. Temperature variation of magnetization for $Fe_{1-x}Li_xV_2O_4$ [where x=0, 0.05, and 0.1 spinels .at H=5000 Oe] and $Fe(Cr_{0.05}V_{0.95})_2O_4$[where x=0 and 0.05 spinels .at H=5000 Oe] and . Inset shows the plot of dM/d T vs.T indicating transitions.

3. Temperature dependence of AC magnetization measured in fields with 100 Hz frequency for $Fe_{1-x}Li_xV_2O_4$ [where x=0, 0.05, and 0.1] around $T_C$. Inset shows the Variation of $T_C$ with respect to $1/R_{V-V}$.

4. The isothermal field dependence of the magnetization at 2 K for $Fe_{1-x}Li_xV_2O_4$ [where x=0 and 0.01] and $Fe(Cr_{0.05}V_{0.95})_2O_4$[where x=0.05].

5. The temperature dependence of ferroelectric polarizaiton of Li and Cr doped FeV2O4

6. (a)The temperature dependences of resistivity for $Fe_{1-x}Li_xV_2O_4$ [where x=0 and 0.01] and $Fe(Cr_{0.05}V_{0.95})_2O_4$[where x=0.05] (b) ln ρ vs 1000/T for $Fe_{1-x}Li_xV_2O_4$ [where x=0 and 0.01] and $Fe(Cr_{0.05}V_{0.95})_2O_4$[where x=0.05].

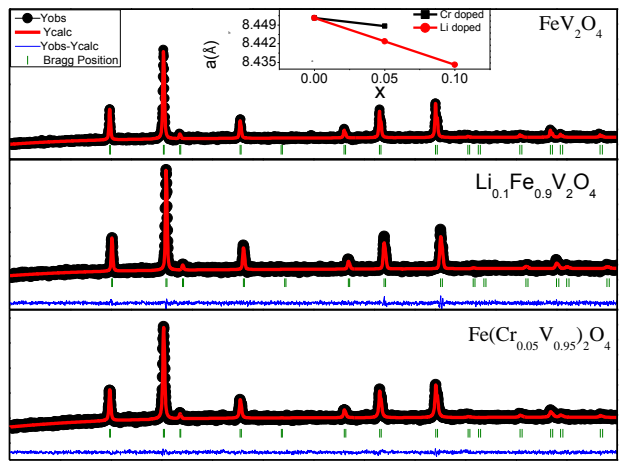

Figure 1

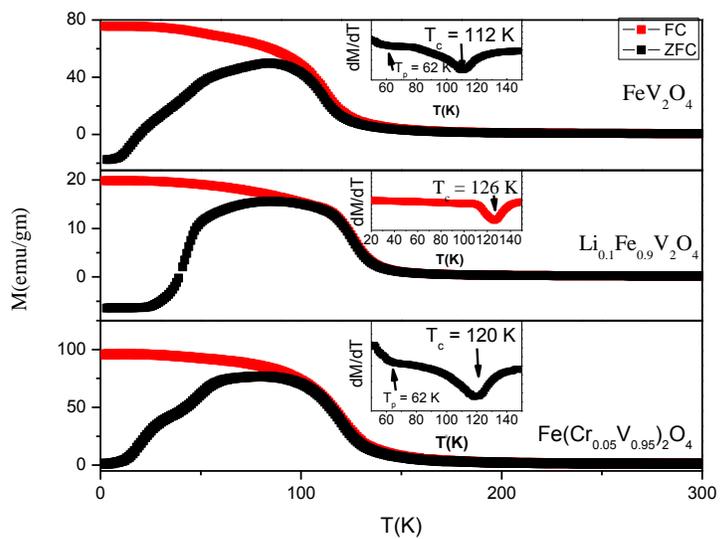

Figure 2

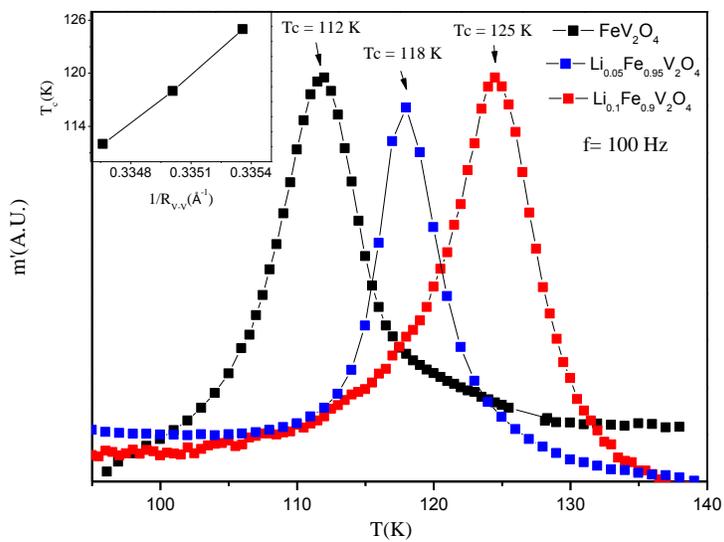

Figure 3

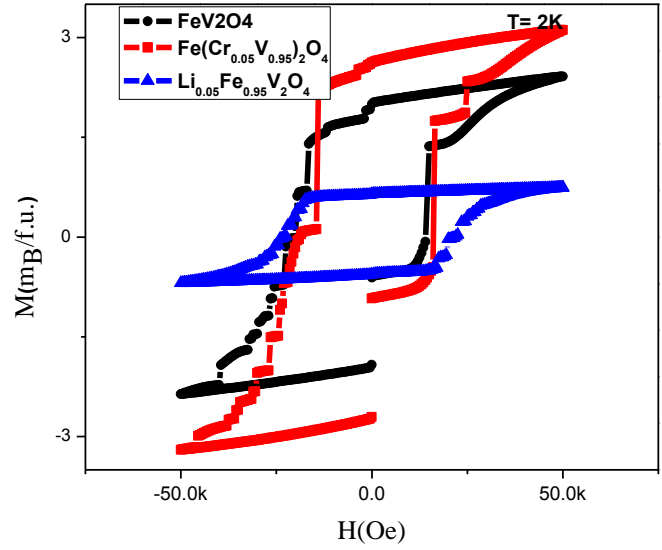

Figure 4

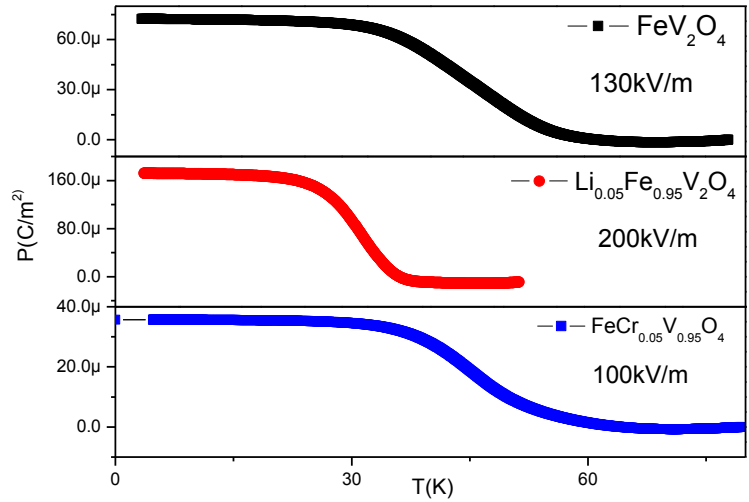

Figure 5

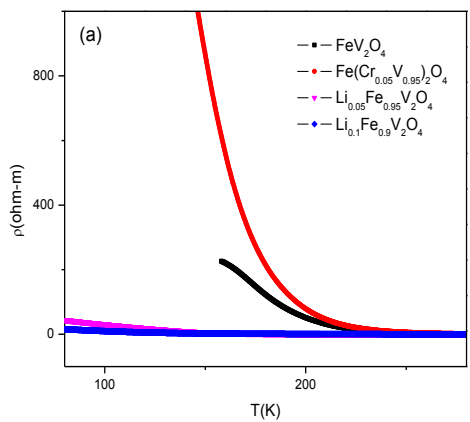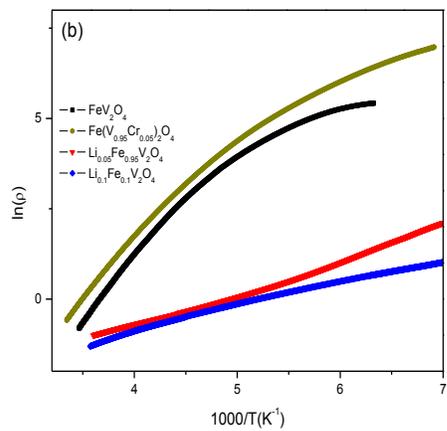

Figure 6